\documentstyle[prl,aps,psfig]{revtex}
\def\be{\begin{equation}}
\def\ee{\end{equation}}
\def\bc{\begin{center}}
\def\ec{\end{center}}

\begin{document}
 
\input epsf.sty
\twocolumn[\hsize\textwidth\columnwidth\hsize\csname %
@twocolumnfalse\endcsname
 
\draft
 
\widetext
\title{Comment on ``Efficient, multiple-range random walk algorithm
to calculate the density of states"}
 
\author{Alfred H\"{u}ller$^a$ and Michel Pleimling$^{a,b}$}
 
\address{
$^a$ Institut f\"ur Theoretische Physik I, Universit\"at Erlangen-N\"urnberg,
D -- 91058 Erlangen, Germany\\
$^b$ Laboratoire de Physique des Mat\'eriaux,$^{**}$
Universit\'e Henri Poincar\'e Nancy I, B.P. 239, \\
F -- 54506 Vand{\oe}uvre l\`es Nancy Cedex, France}
\maketitle
 
\phantom{.}
]
 
\narrowtext

In their recent Letter Wang and Landau\cite{Wan01} have proposed a very
effective MC method of producing evenly distributed histograms for the
degeneracy $g(E_i)$ of the energy levels of different discrete spin models.
The method suffers, however, from 
a drawback. In the course of their calculations Wang and Landau have to reduce their
multiplication factor $f$ to a value $f_{final}$ which differs from 1 only
by $10^{-8}$. At this stage the algorithm comes very close to entropic sampling.
Indeed when $g(E_i)$ has the value $10^8$ then the modified density of states (DOS)
$g(E_i) \longrightarrow g(E_i) * f_{final}$ in the new algorithm and the modified
DOS: $g(E_i) \longrightarrow g(E_i)+1$ in entropic sampling are identical.
Thus the new method has the definite advantage of rapidly building a rough estimator for the
DOS but in the end the work of refining it to useful accuracy is as time consuming as
entropic sampling.

We have ourselves used a similar method as Ref.\ \cite{Wan01} to calculate
histograms of the degeneracy of $g(E_i)$ of three
dimensional Ising models and of the degeneracy
$g(E_0,M_i)$ of 2d-Ising models for a fixed value $E_0$ of
the energy as a function of the magnetization $M_i$. We have also produced
histograms for the three dimensional ANNNI-model.
All these data are unpublished.

In our approach the acceptance probability of a spin flip is also governed by the DOS
which has been accumulated so far. In this respect our approach is identical to that 
of Wang and Landau, but it differs from Ref.\ \cite{Wan01} in two important aspects:
(i) the histogram $g(E_i)$ of the DOS is updated (multiplied by $f$) only
every tenth step,
(ii) the transition observables $T(E_i, \Delta)$ are also evaluated by recording all attempts
to move from an energy $E_i$ to $E_i + \Delta$ with $\Delta = 0, \pm 4, \pm 8$ for the 2d-Ising
model.
On taking the logarithm one obtains the entropy $S(E_i) = \ln g(E_i)$ from the DOS
and the inverse temperature $\beta = \ln \left[ T(E_i + \Delta , - \Delta )/
T(E_i, \Delta ) \right]/\Delta$ from the transition observables. It turns out that the
latter data suffer much less from statistical errors than the former. In contrast to
Ref.\ \cite{Wan01} we do not vary $f$ -- it is kept constant during the simulation.
In the examples which follow we have added $\eta = 0.004$ viz.\ $\eta' = 0.002$ to the
entropy $S(E_i)$ after every 10 attempted spin flips. This leads to multiplication factors
$f = e^\eta$ of 1.004 viz.\ 1.002.

It turns out that the error one makes is roughly proportional to
$\eta$, i.e.,
$\ln g_\eta(E_i) \approx  S(E_i) + \eta f(E_i) + K_\eta$
where $\eta f(E_i)$ is this error and $K_\eta$ is an unimportant constant.
A very good approximation for $S(E_i)$ can therefore be determined from
two MC runs with different values of $\eta$ and $\eta'$:
$S(E_i) \approx \left[  \eta \ln g_{\eta'}(E_i) - \eta' \ln g_\eta(E_i) \right]/
\left( \eta - \eta' \right) + K$.
Fig.\ 1 has been produced for the $32 \times 32$ 2d-Ising model. It shows
the second derivative $d^2s(\varepsilon)/ d \varepsilon^2$ of our simulated data. Here $s = S/N$
and $\varepsilon = E/N$ are the entropy and the energy per spin. These data scatter around
the exact data of Beale\cite{Bea96}. The difference to these data is shown in the inset.
Gaussian smoothing over 20 channels reduces the maximum deviation
to 0.005.

\begin{figure}
\centerline{\epsfxsize=3.00in\epsfbox
{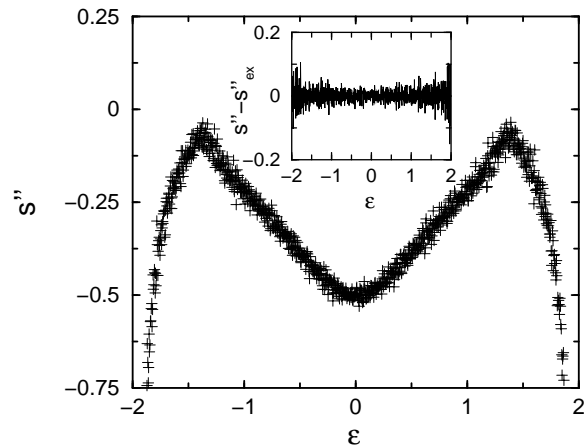}}
\caption{The second derivative $d^2s(\varepsilon)/ d \varepsilon^2$ 
and the difference to the exact data (inset).}
\end{figure} 

As the entropy is determined up to an unimportant constant, the derivatives of the entropy
are more assertive. Therefore one should not compare the entropy itself, but its derivatives with the
exact results\cite{Bea96}. We have chosen the second derivative for this comparison because it enters into
the microcanonically defined specific heat $c(\varepsilon) = - s'^2(\varepsilon)/s''(\varepsilon)$ and it is
thus very important that high quality data are obtained for it.

In summary: Because of our substraction scheme we can keep the entropy increment and concomitantly the
effectivity of the simulation procedure five orders of magnitude above Ref.\ \cite{Wan01}.

We thank Michael Promberger for his advice.


\end{document}